\documentclass[11pt, times, twocolumn]{article}
\usepackage{graphicx}
\usepackage{subfig}
\usepackage{amssymb}
\usepackage{url}
\usepackage{cite}

\begin{document}
\topmargin 0pt
\oddsidemargin 0mm
\renewcommand{\thefootnote}{\fnsymbol{footnote}}
\begin{titlepage}

\vspace{5mm}

\begin{center}
{\Large \bf Strain-tunable energy band parameters of graphene-like GaN} \\

\vspace{6mm} {\large Harihar Behera$^{1, 2}$\footnote{Corresponding author's E-mails: harihar@iopb.res.in;
 behera.hh@gmail.com } and Gautam Mukhopadhyay$^2$\footnote{E-mails: gmukh@phy.iitb.ac.in; g.mukhopa@gmail.com}}    \\
\vspace{5mm}
{\em
$^{1}$ School of Technology, The Glocal University, Mirzapur Pole,
Dist.-Saharanpur, U.P.-247121, India \\
$^2$ Department of Physics, Indian Institute of Technology, Powai, Mumbai-400076, India} \\

\end{center}

\vspace{5mm}
\centerline{\bf {Abstract}}
\vspace{5mm}
We present {\it ab initio} calculations on the effect of in-plane equi-biaxial strain on the structural and electronic
properties of hypothetical graphene-like GaN monolayer (ML-GaN). It was found that ML-GaN got buckled
for compressive strain in excess of 7.281 \%; buckling parameter increased quadratically with compressive strain. The 2D bulk modulus of ML-GaN was found to be smaller than that of graphene and graphene-like ML-BN, which reflects weaker bond in ML-GaN. More importantly, the band gap and effective masses of charge carriers in ML-GaN were found to be tunable by application of in-plane equi-biaxial strain. In particular, when compressive biaxial strain of about 3\% was reached, a transition from indirect to direct band gap-phase occurred with significant change in the value and nature of effective masses of charge carriers; buckling and tensile strain reduced the band gap - the band gap reduced to 50 \% of its unstrained value at 6.36 \% tensile strain and to 0 eV at an extrapolated tensile strain of 12.72 \%, which is well within its predicted ultimate tensile strain limit of 16 \%. These predictions of strain-engineered electronic properties of highly strain sensitive ML-GaN may be exploited in future for potential applications in strain sensors and other nano-devices such as the nano-electromechanical systems (NEMS).\\


\end{titlepage}

\section{Introduction}
Bulk gallium nitride (GaN), which usually adopts a wurtzite (WZ) structure, is a wide direct
 band gap ($\mbox{E}_g(T= 0) = 3.51$ eV) semiconductor, which makes
 it suitable for blue lasers and light emitting diodes (LEDs)
 \cite{1,2}. Its  alloy with indium nitride (InN, $\mbox{E}_g(T = 0) = 0.69$ eV),
  i.e., $\mbox{In}_x\mbox{Ga}_{1-x}\mbox{N}$, has continuously tunable band
   gap that spans the solar energy spectrum by varying $x$ \cite{1, 2}.
   Due to its wide and tunable band gap, high thermal and mechanical
   stability, GaN is one of the most promising materials for applications
   in photovoltaics, optoelectronic devices, high-temperature
   microelectronic devices and solar cells. Currently, band gap
   engineering in bulk GaN is mainly done via doping or alloying,
   for instance with $\mbox{In}$ as in $\mbox{In}_x\mbox{Ga}_{1-x}\mbox{N}$.
    This method suffers from two fundamental problems, viz., (1) when
    In fraction $x$ reaches $0.30$, the quantum efficiency of LEDs
  drops significantly to undesired levels; (2) when $x$ reaches
   $0.32$, a phase separation occurs because of the large atomic
   size mismatch of $\mbox{Ga}$ and In \cite{3, 4}. To overcome
  these problems, alternative methods of band gap tuning in
   GaN through equibiaxial in-plain strains have been proposed
   recently \cite{4} as a method of "strain engineering".
   In the context of GaN based nano-materials, nanowires, nanotubes,
 and nanospirals of GaN have been synthesized which showed great potential
 for fabricating wide-spectrum LEDs and other nano-scale devices \cite{5, 6,7}. In particular, a density functional theory (DFT) study in 2006 \cite{8} predicted that when the layer number of $(0001)$-oriented WZ materials (e.g., AlN, BeO, GaN, SiC, ZnO, and ZnS) is small, the wurtzite structures transform into a new form of stable hexagonal graphite-like or hexagonal boron nitride (BN)-like structure. This prediction has been confirmed by experiments in
 respect of ZnO in 2007 \cite{9}. Recently the thickness range of graphitic WZ
 films (e.g., AlN, GaN, InN, SiC, BeO, ZnO) are theoretically shown to depend on strain and can be extended to much thicker films by epitaxial strain \cite{10}. In particular for GaN film, it has been shown \cite{10} that (i) a 10-atomic-layer $(0001)/(000\overline{1})$ GaN film transforms into a more stable 5 monolayer (ML) graphitic film having no destabilizing dipole moments in its surface, (ii) For GaN films beyond 10-layers, the WZ structure tends to be stable due to the charge transfer from the $(000\overline{1})$ bilayer to the $(0001)$ bilayer, counteracting the intrinsic dipole perpendicular to the film, (iii) with $10\%$ biaxial tensile strain, the graphitic phase of GaN was found to be stable up to 24 layers and (iv) the band gap of the stable strained graphitic films are also shown to be tunable over a wide range either above or below that of the bulk WZ phase.\\
 \indent
The study of 2D crystals is an emerging field of research inspired by the recent phenomenal growth in the research on graphene
(a one atom-thick nanocrystal of carbon atoms tightly-bound by the sp$^2$-hybridized bonds in a 2D hexagonal lattice) which promises many novel applications \cite{11, 12, 13, 14}. Representative samples of some 2D/quasi-2D nanocrystals that have been synthesized recently include ZnO \cite{9}, BN \cite{11, 12, 13, 14}, MoS$_2$ \cite{11, 12, 13, 14}, MoSe$_2$ \cite{11, 12, 13, 14}, Bi$_2$Te$_3$ \cite{11, 12, 13, 14}, Si \cite{15}. With rapid advancement of synthesis techniques, the synthesis of GaN monolayer (ML-GaN) is expected in near future. Therefore, understanding and controlling the electronic
properties of ML-GaN is of much significance both from fundamental and application point of view. Although several theoretical studies have been reported \cite{16, 17, 18, 19, 20, 21} to this end, the effect
 of strain on the structural and the electronic properties of ML-GaN remains unclear. Since strain-engineering is an important method to tailor the structural and electronic properties of nanomaterials \cite{22, 23, 24}, using DFT we have investigated the effect of equi-biaxial inplane strain on ML-GaN. Here we report the results our study which simulates an ideal experimental situation in which ML-GaN is supported on a flat stretchable substrate.
\begin{figure}
 \includegraphics[scale=0.9]{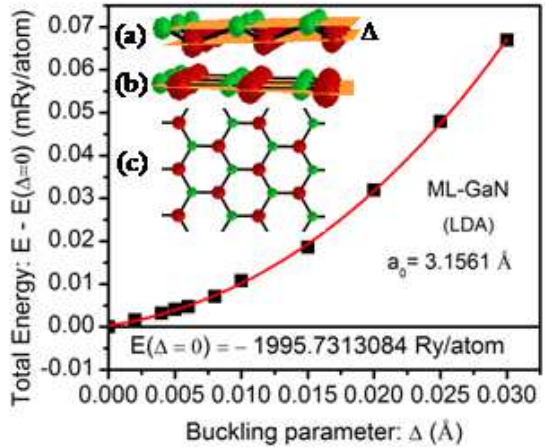}
 \caption{Buckling probe of planar (PL) ML-GaN based on the energy
 minimization procedure with two atoms per unit supercell. Insets show the ball-stick model for (a) the side
  view of buckled (BL) ML-GaN, (b) the side view of planar ML-GaN and
  (c) the top-down view of both PL and BL ML-GaN. In BL ML-GaN, Ga atoms
  (large balls, red) and N atoms (small balls, green) are in two different
   parallel planes; buckling parameter is the perpendicular distance
   between those parallel planes and $\Delta = 0 $\AA \, for PL ML-GaN. }
\end{figure}

\section{Computational Methods}
The calculations have been performed by using the density functional theory (DFT) based full-potential (linearized) augmented plane wave plus local orbital (FP-(L)APW+lo) method \cite{25}, which is a descendant of FP-LAPW method \cite{26}. We use the elk-code \cite{27} and the Perdew-Zunger variant of local density approximation (LDA) \cite{28}, the accuracy of which have been successfully validated in our previous studies \cite{29, 30, 31, 32, 33, 34, 35, 36} of graphene and some
 graphene-like 2D nano-crystals. For plane-wave expansion in the interstitial
 region, we have chosen $|{\bf G}+{\bf k}|_{\mbox{max}}\times{\mbox{R}}_{\mbox{mt}}= 8.5$ (R$_{mt}$
 is the smallest muffin-tin radius in the unit cell) for deciding the
 plane-wave cut-off. The Monkhorst-Pack \cite{37} {\bf k}-point grid size
 of $30\times30\times1$ was chosen for all calculations. The total energy was converged within $2\mu$eV/atom. We simulate the 2D-hexagonal structure of ML-GaN as a 3D-hexagonal supercell with a large value of $c$-parameter ($= |{\bf c}| = 40$ a.u.) and two atoms per unit supercell. The application of in-plane equi-biaxial strains about $\pm 10\%$ was simulated by varying the in-plane lattice parameter $a (=|a| = |b|); \epsilon = [(a - a_0)/a_0]\times 100$, where $a_0$ is the ground state in-plane lattice constant of ML-GaN. Our consideration of such large values ($\pm 10\%$) of strain is inspired by
 (i) the first principles study of mechanical properties of ML-GaN \cite{21}, where ML-GaN is shown to have an ultimate strain value of 0.16 under biaxial deformation,
 (ii) the recent theoretical studies on strained GaN films, where strains in the range of ($-5\% \mbox{ to} +10\%$)\cite{10} and ($-20\%\mbox{ to} +20\%$)\cite{4} have been considered and (iii) the fact that graphene (whose honeycomb structure the ML-GaN adopts) is experimentally demonstrated \cite{38} to sustain in-plane tensile elastic strain in excess of $20\%$, although we do not yet clearly comprehend the maximum compressive strain graphene can sustain while preserving its planar structure. However, for the sake of symmetry in the compression variation, and as a matter of curiosity, we have considered compressive strains up to $10\%$. For the structural stability of ML-GaN, we have considered two possible structures of ML-GaN (with two atoms per unit cell) as presented in Figure 1.


\section{Results and Discussions}
\subsection{Structural and Mechanical Properties}
For an assumed flat ML-GaN, our calculated LDA value of $a_0 = 3.1561$ \AA\, corresponds to the Ga-N bond length $d_{\mbox{Ga-N}} = a_0/\sqrt{3} = 1.822$ \,\AA, which is close to the reported LDA value of $d_{\mbox{Ga-N}} = 1.85$ \,\AA\, by \c{S}ahin et al. \cite{16}, GGA values of 1.87 \AA \, by Chen et al. \cite{17}, 1.83 \AA \, by Xiao et al. \cite {18}, 1.852 \AA \, by Peng et al. \cite {21}. It is well known that usually LDA underestimates and GGA overestimates the lattice constant. In our previous studies of graphene and silicene \cite{29} and germanene \cite{30}, we have demonstrated that the value of $c-$parameter chosen in the construction of supercell for simulation of 2D hexagonal structures also affects the value of in-plane lattice constant $a_0$: larger value of $c-$parameter yields a smaller value of $a_0$. Since we use a different method and a different value of $c-$parameter, the slight disagreement of our result on $a_0$ with other theoretical results is well understood and acceptable. Our assumption on the flat ML-GaN structure was tested correct by the calculated results depicted in Figure 1, based on the principle of minimum energy for the stable structure. This corroborates the results
of previous theoretical studies \cite{8, 9, 10, 16, 17, 18, 21} based on different methods and the stability of flat ML-GaN is attributed to the strong in-plane sp$^2$ hybridized bonds between the Ga and N atoms. \\
  \begin{figure}
 \centering
 \subfloat[For $\epsilon = -7\%.$ ]{
 \label{fig:subfig:a} 
 \includegraphics[scale=0.5]{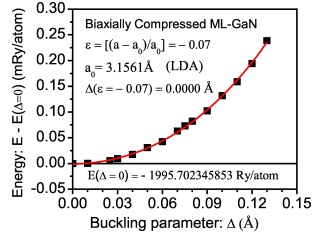}}
 \hspace{0.1in}
\subfloat[For $\epsilon = - 8\%.$ ]{
 \label{fig:subfig:b} 
 \includegraphics[scale=0.5]{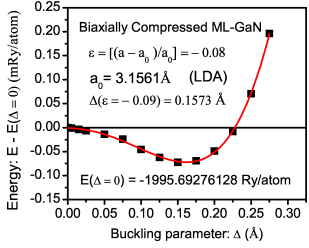}}\\
  \vspace{0.02in}
 \subfloat[For $\epsilon = - 9\%.$ ]{
 \label{fig:subfig:c} 
 \includegraphics[scale=0.5]{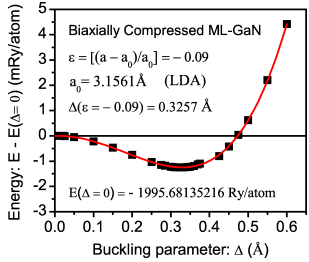}}
  \hspace{0.1in}
 \subfloat[For $\epsilon = - 10\%.$ ]{
 \label{fig:subfig:d} 
 \includegraphics[scale=0.5]{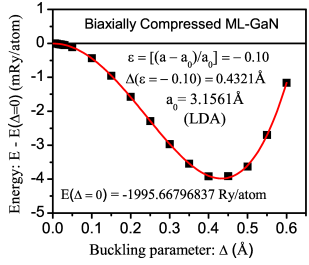}}
\caption{Buckling probe of ML-GaN at four
different values of compressive strain of (a) 7\%, (b) 8\%, (c) 9\% and (d) 10\%.}
 \label{fig:subfig} 
\end{figure}
 \indent
The calculated 2D bulk modulus (also called in-plane Young's modulus) of ML-GaN, $B_{\mbox{2D}} = A(\partial ^{2}E/\partial A^2)|_{A_{min}} = 109.45$ N/m, (where  $A$ is the area of the periodic cell of the 2D lattice and $A_{min}$ is the area with minimum energy) is in good agreement with reported results: 110 N/m \cite{16}, 109.4 N/m \cite{21}. This value is smaller than the corresponding calculated values of graphene (214.41 N/m) and ML-BN (181.91 N/m) \cite{39}, reflecting the weaker bonds in ML-GaN than those in graphene and ML-BN. Therefore, ML-GaN is expected to be useful in applications requiring materials with less tougher behavior than graphene and ML-BN. Further, since $B_{\mbox{2D}}\mbox{(ML-GaN)} \approx \frac{1}{2}B_{\mbox{2D}}\mbox{(Graphene)}$ as seen above, we expect ML-GaN to sustain elastic tensile strains $\approx 10\%$ considering the fact that graphene sustains elastic tensile strains in excess of 20\% \cite{38}.
However, to know the compressive strain value that ML-GaN can sustain without change in its planar structure, we have made some study by probing the buckling in ML-GaN at different values of compressive strains. The results of this probe for four values of compressive strains, viz., $\epsilon = - 7\%, -8\%, -9\% \,\mbox{and} -10\%$, are shown in Figure 2, in which it is seen that ML-GaN sustains compressive strain of 7\% without buckling and with higher compressive strain values of 8\%, 9\% and 10\%, buckling parameter assumes nonzero and higher values. From these data we inferred that there is a critical compressive strain value beyond which buckling starts. To determine this critical point, we have plotted the buckling parameters calculated at six compressive strains of 7.5\%, 8.0\%, 8.5\%, 9.0\%, 9.5\% and 10\% in Figure 3, where our data fit well with the following second order polynomial, viz.,
  \begin{equation}
  \Delta (|\epsilon|) = - 3.42492 + 0.6961 |\epsilon| - 0.03104 |\epsilon|^2
  \end{equation}
where $\Delta$ is in \AA \, and $|\epsilon|$ represents the absolute value of compressive strain in \%. One of the solutions of Eq.(1) for $\Delta(|\epsilon|) = 0$ \AA \, is $|\epsilon| = 7.281\%$ which is the critical compressive strain beyond which ML-GaN becomes buckled. This is in agreement with our calculated result of Figure 2(a), where we found no buckling at compressive strain of 7\%. Eq.(1), which is valid for compressive strain value $|\epsilon| \leq 10\%$, may be useful for calculating the buckling parameter corresponding to a particular value of compressive strain $|\epsilon| \leq 10\%$, which in turn may be used to calculate the electronic properties of ML-GaN under compressive strains $10\% < |\epsilon| \leq 7.281\%$. So we have used this approach to calculate the electronic properties of buckled ML-GaN under high compressive strain values $|\epsilon| \leq 10\%$, which we present in the next section.
\begin{figure}
 \centering
 \includegraphics[scale=0.9]{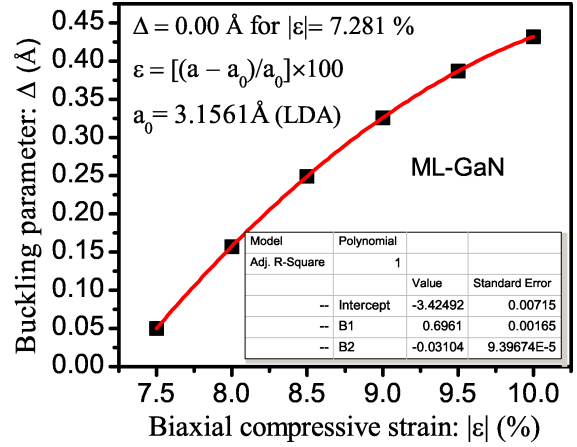}
 \caption{Variation of buckling parameter of ML-GaN with absolute value of compressive strain.}
\end{figure}
\begin{figure}
 \centering
 \includegraphics[scale=0.85]{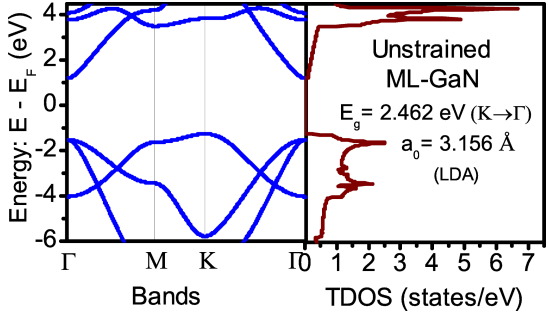}
 \caption{Bands and total DOS of unstrained ML-GaN within LDA.
 $\mbox{E}_F$ is the Fermi energy.}
 \end{figure}

\subsection{Electronic Properties}
The electronic band structures and total density of states (TDOS) plots of unstrained ML-GaN are depicted in Figure 4. As seen in Figure 4, unstrained ML-GaN is an indirect band gap (E$_{g} = 2.462$ eV ($K\rightarrow\Gamma$), LDA value) semiconductor with conduction band minimum (CBM) located at the $\Gamma$ point and valence band maximum (VBM) located at the $K$ point of the hexagonal Brillouin Zone (BZ). However, the actual band gap is surely larger than this since LDA is well known to underestimate the gap. Our calculated LDA band gap of 2.462 eV may be compared with reported LDA value of 2.27 eV by \c{S}ahin et al. \cite{16}, GGA values of 1.95 \AA \, by Chen  et al. \cite{17}, 2.46 eV by Xiao et al. \cite{18}. Using the GW approximations, the GW band gap of ML-GaN have been estimated at 5.00 eV (LDA+GW$_0$) by \c{S}ahin et al. \cite{16} and 4.14 eV (GGA+GW) by Chen et al. \cite{17}, which showed that ML-GaN is a wide indirect band gap semiconductor. Although both LDA and GGA do not yield the band gap correctly, these are powerful enough to predict a correct trend in variation of the gap \cite{4, 17, 40, 41}. Since we focus on the trend  as well as the relative variation of E$_g$ rather than its absolute value, we employed the computationally simpler and less time-consuming LDA for this study.\\
\begin{figure}
 \centering
 \includegraphics[scale=0.9]{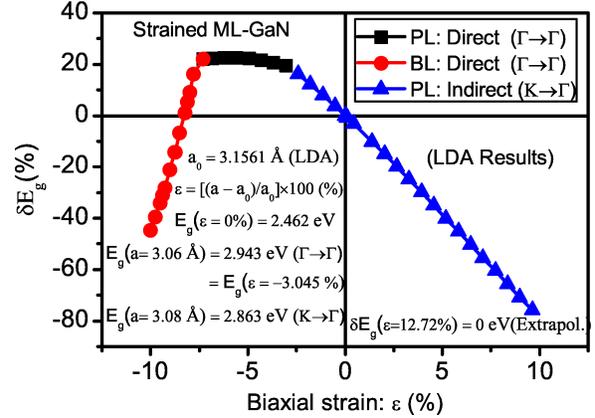}
\caption{Relative variation in band gap $\delta \mbox{E}_g$ (with respect to the unstrained
 band gap $\mbox{E}_g(0) = 2.462$ eV) with in-plane homogeneous biaxial strain $\epsilon$.}
\end{figure}

We simulated the application of in-plane equi-biaxial strain (i.e., hydrostatic strain) to ML-GaN by varying the in-plane lattice parameter $a$ from 2.84 \AA \, to 3.46 \AA \, with small increments of 0.02 \AA. With our
calculated value of $a_0 = 3.1561$ \AA \, for unstrained ML-GaN, we thus simulated the strains in the range ($-10\% \,\mbox{to}\, +9.63\%$). Our calculated results on the in-plane equi-biaxial strain-induced modifications of the band gap of ML-GaN are depicted in Figure 5, which shows both the trend as well as the relative change in E$_g$ defined by
\begin{equation}
\delta\mbox{E}_g = [(\mbox{E}_g(\epsilon) - \mbox{E}_g(0))/\mbox{E}_g(0)]\times 100 \hspace{0.5cm} \,(\%)
\end{equation}
\noindent
with respect to the unstrained band gap $\mbox{E}_g(0) = 2.462$ eV (the relative change is expected to be more meaningful and free from computational method dependent errors). For 3.08 \AA $\leq a \leq  3.46$ \AA\, ($-3.045 \% \leq \epsilon \leq  9.629 \%$), E$_g$ remains indirect
($K\rightarrow\Gamma$) and its variation with strain is linear; for 2.84 \AA $\leq   a \leq  3.06$ \AA \, $(- 10\% \leq \epsilon < 3.045 \%)$, E$_g$ remains direct ($\Gamma \rightarrow \Gamma$) and its variation is non-linear with strain in the planar phase and almost linearly decreasing in the buckled phase. The band gap variation is asymmetric with respect to $\epsilon$ not only in the trend but also in the character (transition from indirect to direct band gap). For tensile strains the relative change in E$_g$ is very much pronounced; E$_g$ dramatically drops to $50\%$ ($\delta \mbox{E}_g =-50\%$) of its unstrained value at $\epsilon = 6.36\%$ (obtained by linear interpolation) and to $0$ eV ($\delta \mbox{E}_g = -100\%$) at $\epsilon=12.72\%$ (obtained by linear extrapolation).
In contrast, for high compressive strain, the reduction in E$_g$ is about $44.78\%$ at $\epsilon = - 10\%$.
More interestingly, a transition from indirect to direct band gap occurs
at about $3\%$ compressive strain as noted in Figure 5 for E$_g(a = 3.08 \mbox{\AA}) = 2.863$ eV
($K \rightarrow \Gamma$) and E$_g (a = 3.06 \mbox{\AA})
= 2.943 $ eV ($\Gamma \rightarrow \Gamma$). Further, the direct band gap (which is
larger than pristine indirect band gap) non-linearly saturates ($\delta \mbox{E}_g = +22\%$) with increasing
compressive strain until ML-GaN assumes buckling, which in turn induces a reduction in E$_g$ as seen in Figure 5.
This property of ML-GaN may be exploited for non-linear high-frequency
optical applications in future. It is to be noted that, except for the
occurrence of the direct to indirect gap-phase transition at about $3\%$
compressive strain and the positions of CBM and VBM, the present trend in
the variation of E$_g$ with $\epsilon$ is similar to those recently predicted for
bi-axially strained bilayer GaN (2-ML-GaN)\cite{10} and ultrathin GaN
films\cite{4} in graphitic phase; the positions of VBM and CBM in 2-ML-GaN are
not reported in ref. \cite{10},  but in graphitic GaN films, the VBM is at
H point and CBM is positioned at $\Gamma$ point of the 3D hexagonal BZ \cite {4}.
Thus the electronic properties of ML-GaN, are different from those of few layer
graphitic GaN films. This is not surprising, as we know that the electronic
properties of monolayer, bilayer and few-layer graphene are all different.
A related theoretical study \cite{42} of bi-axially strained direct band
gap ML-BN (corroborated by our recent study \cite{31}), showed strain
induced gap-phase transition from direct to indirect band gap at
$\epsilon = -1.53\%$, and E$_g$ follows linear variation for strains below
and above $\epsilon = -1.53\%$, which substantiate our confidence on the
present results. Since the band gap of ML-GaN is very sensitive to strain,
it is a potential candidate for new-generation strain sensor.\\
\begin{figure}[h]
 \centering
 \includegraphics[scale=1.0]{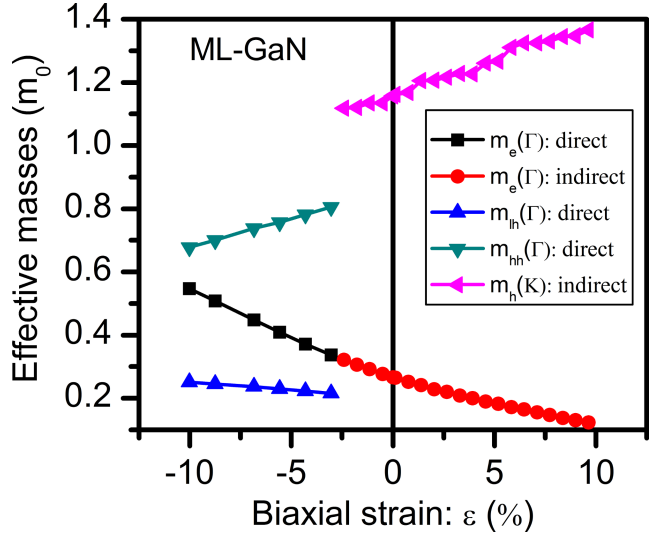}
 \caption{Variation of effective masses of charge carriers in ML-GaN with in-plane homogeneous biaxial strain $\epsilon$ (\%).}
\end{figure}
The nature of and trend in the strain-induced variation of effective masses
 of the electrons $m_e(\Gamma )$ for both indirect as well as direct gap
 phases, holes $m_h(K)$ for the indirect gap phase, light holes
 $m_{lh}(\Gamma)$ and heavy holes $m_{hh}(\Gamma)$ for direct gap-phase,
 calculated at the band edges at the special points $K$ and $\Gamma$ as
 appropriate for the two gap-phases, are shown in Figure 6. For unstrained
 ML-GaN, we calculated $m_e(\Gamma ) = 0.266m_0$ and $m_h(K) = 1.157m_0$ ($m_0$
 is electron's rest mass), which are larger than the band edge electron effective
  mass of $0.20m_0$ for the bulk wurtzite GaN \cite{1, 2}. As seen in Figure 6,
  unlike the case with electrons, the character and the value of
  effective masses of holes abruptly change at about $\epsilon = - 3\%$.
  The discontinuity in the variation of the effective masses of holes
   is due the change of the gap-phase which makes the curvature of the
  VBM at $K$ point in the indirect gap-phase much smaller than the
 curvature of VBM at the $\Gamma$ point in the direct gap-phase.
 Since the mobility of electrons and holes depend on their effective masses,
 the carrier mobility in ML-GaN can also be controlled by strain engineering.


 \section{Conclusion}
 In summary, we have theoretically explored the effects of in-plane equi-biaxial strains on the
  structural and electronic properties of ML-GaN in graphene-like planar honeycomb structure
  employing DFT based all electron full-potential calculations.
 In this study we found that ML-GaN gets buckled for compressive strain in excess of 7.281 \%.
 Within our considered compressive strain limit of 10\%, buckling parameter varies quadratically with
 compressive strain with an increasing trend with increasing compression. The calculated
 value of 2D bulk modulus of ML-GaN reflects weaker bond in ML-GaN in comparison to
 that in graphene and ML-BN. More importantly, our calculations show that the band
 gap and effective masses of charge carriers in ML-GaN can be significantly tuned
 and manipulated by application of in-plane equi-biaxial strain.
 In particular, when compressive biaxial strain of about $3\%$ is reached,
 a transition from indirect to direct band gap-phase is predicted to occur
 with significant change in the value and nature of effective masses of charge
 carriers; the band gap reduces to $50\%$ of its unstrained value at $6.36\%$
 tensile strain and to $0$ eV at an extrapolated tensile strain of $12.72\%$.
 These interesting predictions of strain-engineered electronic properties of
 highly strain sensitive ML-GaN may be exploited in future for novel device
 applications such as nano-mechatronics, strain sensor, nano-electromechanical systems
 (NEMS) and nano-opto-mechanical systems (NOMS).\\

\section*{Acknowledgments}
The preliminary results of this work were presented as poster in the ``Graphene Week 2014"
conference held during 23-27 June 2014, at Chalmers University of Technology, Gothenberg, Sweden. The
High Performance Computing Facility (Nebula-Cluster) of IIT Bombay was used for all the calculations in this paper.

\bibliographystyle{elsarticle-num}

\end{document}